\documentclass[twocolumn,secnumarabic,amssymb, nobibnotes, aps, prb, showpacs]{revtex4}
\usepackage[utf8]{inputenc}
\usepackage[english]{babel}
\usepackage{amsmath}
\usepackage{amsfonts}
\usepackage{amssymb}
\usepackage{makeidx}
\usepackage{graphicx}
\usepackage{lmodern}
\usepackage{epstopdf}

\begin{document}
\title{Transmission-phase of an electron in a quantum point contact}
\date{\today}
\author{B.G.C. Lackenby}
\author{O.P. Sushkov}
\affiliation{School of Physics, University of New South Wales, Sydney 2052, Australia}
\pacs{73.23.Ad, 73.40.Lq, 73.63Rt, 73.21.Hb}
\begin{abstract}
For the first time we calculate the electron transmission phase through a 
quantum point contact (QPC). The QPC is considered in the saddle point
approximation in the single-electron picture.
We show that when the electron energy
is close to the height of the  potential barrier
the transmission-phase depends linearly on the energy.
The coefficient in the linear dependence is logarithmically enhanced
by the ratio of the  height over the curvature of the barrier potential.
We compare the calculated transmission phase with the first experimental measurements of the
phase.
\end{abstract}
\maketitle

The conductance of a quantum point contact (QPC) - a one dimensional constriction in a 
two dimensional electron gas - has been known to be quantized in units of $G_0=2 e^2 /
h$ since 1988 \cite{wharam88,vanwees88}. The observed conductance plateaus 
can be understood in the single-electron picture and the saddle point
potential model of the QPC~\cite{land,buttiker90}.

The transmission probability of a saddle point potential was first
calculated in Ref.~\cite{buttiker90}. The transmission probability
describes conductance of a QPC. However, to the best of our knowledge
the transmission phase has never been calculated. This may be because the transmission phase is much more difficult to experimentally measure compared to the transmission probability. However the QPC transmission phase has been recently measured for the first time~\cite{kobayashi13} and therefore it is important to provide a theoretical basis for this property as more experiments are performed and the field expands.
In the present work we calculate the transmission phase and compare it
with experiment~\cite{kobayashi13}.
The linear relationship of the transmission phase in Ref.~\cite{kobayashi13} is attributed to many-body effects however in this paper we show that, amazingly, the linear relationship is logarithmically robust and only a consequence of the saddle point potential in a single electron model. Although it is not the focus of this paper, we will also mention how we expect electron correlations to affect the transmission phase in the 0.7 regime.

We consider the single-electron picture of the QPC within the saddle point
potential approximation. Essentially we use the same approach as Buttiker in his seminal paper~\cite{buttiker90}, this approach is certainly valid for higher conductance steps and according to Refs.~\cite{sloggett98,lunde,bauer13} 
at zero temperature and at zero potential bias the single electron approach
is justified  even for the lowest conductance step since  inelastic 
scattering channels are closed.

\begin{figure}[htbp]
\includegraphics[scale=0.4]{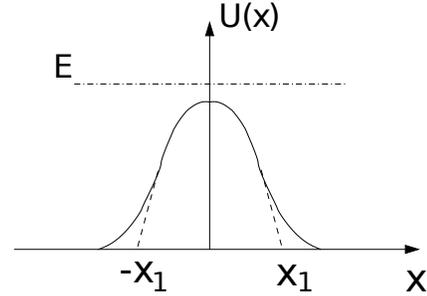}
\caption{The effective potential is sketched by the solid line.
The horizontal dashed-dotted line indicates the energy level.
The dashed line shows the parabolic approximation to the potential, $U(x)$.
}
\label{fig1}
\end{figure}

We assume that far from the QPC the potential is zero,
this is the reference level. Near the QPC the potential
has a saddle point shape
\begin{equation}
\label{spp}
V = V(0) - \dfrac{1}{2}m\omega_x^2x^2 + \dfrac{1}{2}m\omega_y^2y^2 \ ,
\end{equation}
where $m$ is effective mass of the electron.
The electric current flows in the x-direction.
In an adiabatic approximation the variables in the two-dimensional Schrodinger equation
are separated and the transmission problem is reduced to the solution of one dimensional Schrodinger 
equation with an effective potential $U(x)$~\cite{buttiker90}
\begin{equation}
\label{Hamiltonian}
\left(\frac{p_x^2}{2 m} + U(x) \right)\psi(x)=E\psi(x) \ .
\end{equation}
The potential is peaked at $x=0$ and in the vicinity of this point the potential is
\begin{eqnarray}
\label{u0}
&&U(x)\approx U_0-\frac{1}{2}m\omega_x^2x^2\\
&&U_0=V(0)+\hbar\omega_y(n+1/2) \nonumber ,
\end{eqnarray}
where $n=0,1,2,3...$ indicates a  transverse channel.
The potential of a QPC is sketched in Fig.~\ref{fig1} by a solid line and the parabolic approximation is the dashed line.
\newline
The parabolic approximation (\ref{u0}) deviates from the real potential
at large distances, the approximation is valid at $|x|\ll x_1$,
see Fig.~\ref{fig1}.
Within the parabolic region the electron wave function is proportional
to the function of the parabolic cylinder $D_{\nu}$~\cite{BM}
\begin{eqnarray}
\label{dn}
&&\psi(x) \propto D_{\nu}(\sqrt{2}\xi e^{-i\pi/4})\nonumber\\
&&\xi=x/x_0\nonumber\\
&&x_0=\sqrt{\frac{\hbar}{m\omega_x}}\nonumber\\
&&\nu=-\frac{1}{2}+i\epsilon\nonumber\\
&&\epsilon=\frac{E-U_0}{\hbar\omega_x} \ .
\end{eqnarray}
Here we assume that the electron is incident from the left.
Asymptotically, $x_1 \gg |x| \gg x_0$, the wave function (\ref{dn})
consists of the incident (I), reflected (R), and transmitted (T) waves~\cite{BM}
\begin{eqnarray}
\label{asimpt1} 
&& x <0, \ \ \psi=\psi_I+\psi_R:\nonumber\\
&&\psi_I = \frac{\sqrt{2 \pi} e^{-\pi\epsilon/4}}
{\Gamma\left(\tfrac{1}{2} - i\epsilon\right)}
\frac{1}{\sqrt{2|\xi|}} \exp\left(\dfrac{-i\xi^2}{2} - 
\dfrac{i \epsilon}{2}\ln\left(2\xi^2\right) + \dfrac{i \pi}{8}\right)\nonumber \\
&&\psi_R = e^{-3\pi\epsilon/4}\frac{1}{\sqrt{2|\xi|}}
\exp\left(\dfrac{i\xi^2}{2} + \dfrac{i \epsilon}{2}\ln\left(2\xi^2\right) + \dfrac{3i \pi}{8}\right)\nonumber\\
&& x > 0, \ \ \psi=\psi_T:\nonumber\\
&&\psi_T = e^{\pi\epsilon/4}\frac{1}{\sqrt{2|\xi|}}
\exp\left(\dfrac{i\xi^2}{2} + \dfrac{i \epsilon}{2}\ln\left(2\xi^2\right) + \dfrac{i \pi}{8}\right).
\end{eqnarray}
It is important to note that the phases in these wave functions (\ref{asimpt1}) contain
the logarithmic term, $\dfrac{i \epsilon}{2}\ln\left(2\xi^2\right)$, somewhat similar to
the logarithmic phase in the Rutherford scattering  from a Coulomb field~\cite{LL}.
The logarithmic term is the origin of the logarithmic enhancement of the linear energy
dependence in the transmission phase, see below.
The wave functions (\ref{asimpt1}) immediately give the well known transmission
probability of the parabolic barrier~\cite{BM}
\begin{eqnarray}
\label{tr}
T=\frac{e^{\pi\epsilon}}{2\pi}
\Gamma\left(\tfrac{1}{2} - i\epsilon\right)
\Gamma\left(\tfrac{1}{2} + i\epsilon\right)
=\frac{1}{1+e^{-2\pi\epsilon}}.
\end{eqnarray}
Conductance of the QPC  is $G=G_0T$, 
where the transmission coefficient is taken at energy equal to the Fermi energy,
see Ref.~\cite{buttiker90}.

While the wave function (\ref{dn}),(\ref{asimpt1}) is sufficient to determine the 
transmission probability, it is not sufficient to determine the transmission phase.
The point is that the transmission phase is defined at $x \gg x_1$, while the wave
function (\ref{dn}),(\ref{asimpt1}) is valid only at $|x| \ll x_1$, see Fig. \ref{fig1}.
We use semiclassical approximation to propagate (\ref{asimpt1}) to the distances
$|x| \gg x_1$. Let us first do the transmitted wave.
At any $x \gg x_0$ the transmitted wave function can be represented as
\begin{eqnarray}
\label{pt1}
&&\psi_T(x) \propto e^{i\phi_T}e^{{\tfrac{i}{\hbar}}\int_0^{x}  p(x) dx}\nonumber\\
&& p(x)=\sqrt{2m[E-U(x)]} \ ,
\end{eqnarray} 
where $\phi_T$ is a phase. 
To determine $\phi_T$ we calculate (\ref{pt1}) at $x_0 \ll x \ll x_1$. 
A straightforward integration gives
\begin{eqnarray}
\label{pt2}
\psi_T(x) \propto e^{i\phi_T}
\exp\left(\tfrac{i\xi^2}{2}
+\tfrac{i\epsilon}{2}\ln\left(2\xi^2\right)
- \tfrac{i\epsilon}{2}\ln \epsilon
+\tfrac{i\epsilon}{2}
\right).
\end{eqnarray} 
Comparing this with $\psi_T$ from  (\ref{asimpt1}) we find $\phi_T$,
\begin{equation}
\label{ft}
\phi_T = \dfrac{\epsilon}{2}\left(\ln\epsilon - 1\right) + \dfrac{\pi}{8} \ .
\end{equation}
Now, using (\ref{pt1}) we can calculate $\psi_T$ at $x \gg x_1$.
The integral $\int_0^{x}  p(x) dx$ depends on the exact shape of the potential
at $x \sim x_1$. However, the shape dependent contribution to the phase is almost independent
of the electron energy while we are interested only in the energy dependent part of 
the phase.
Therefore, we assume that the parabolic approximation (\ref{u0}) is valid up
to $x=x_1$ and we also assume that $U(x)=0$ at $x > x_1$, see the dashed line in
Fig. \ref{fig1}. We will discuss this assumption again later.
Using (\ref{pt1}) and (\ref{ft}) we find the transmitted wavefunction at $x> x_1$
\begin{eqnarray}
\label{pt3}
&&\psi_T \propto e^{i\delta_T+ik_0x}\nonumber\\
&&\delta_T=\frac{\epsilon}{2}\ln\left(4{\cal U}\right)
+{\cal U}+\frac{\pi}{8}-ik_0x_1\ .
\end{eqnarray}
Here $k_0=\sqrt{2mE}/\hbar$ and ${\cal U}=\frac{U_0}{\hbar\omega_x}$.

Now we perform a similar calculation for the incident wave,
note that $x$ is negative, $x=-|x|$.
In semiclassical approximation the incident wave is
\begin{eqnarray}
\label{pI1}
\psi_I(x) \propto e^{i\phi_I}e^{-{\tfrac{i}{\hbar}}\int_0^{|x|}  p(x) dx}\ ,
\end{eqnarray} 
where $\phi_I$ is a phase. 
To determine $\phi_I$ we calculate (\ref{pI1}) at $x_0 \ll |x| \ll x_1$. 
A straightforward integration gives
\begin{eqnarray}
\label{pI2}
\psi_I(x) \propto e^{i\phi_I}
\exp\left(\tfrac{-i\xi^2}{2}
-\tfrac{i\epsilon}{2}\ln\left(2\xi^2\right)
+ \tfrac{i\epsilon}{2}\ln \epsilon
-\tfrac{i\epsilon}{2}
\right).
\end{eqnarray} 
Comparing this with $\psi_I$ from  (\ref{asimpt1}) we find $\phi_I$,
\begin{equation}
\label{fi}
\phi_I = -\dfrac{\epsilon}{2}\left(\ln\epsilon - 1\right) + \dfrac{\pi}{8}
+arg\left[\Gamma\left(\tfrac{1}{2} + i\epsilon\right)\right]
 \ .
\end{equation}
Now, using (\ref{pI1}) we can calculate $\psi_I$ at $x < -x_1$.
Again, we assume that the parabolic approximation (\ref{u0}) is valid
at $-x_1< x < 0$ and  $U(x)=0$ at $x < -x_1$.
Using (\ref{pI1}) and (\ref{fi}) we find the incident wavefunction at $x < -x_1$
\begin{eqnarray}
\label{pI3}
&&\psi_I \propto e^{i\delta_I+ik_0x}\\
&&\delta_I=-\frac{\epsilon}{2}\ln\left(4{\cal U}\right)
-{\cal U}+\frac{\pi}{8}+
arg\left[\Gamma\left(\tfrac{1}{2}+ i\epsilon\right)\right]
+ik_0x_1\ .\nonumber
\end{eqnarray}
Having (\ref{pt3}) and (\ref{pI3}) we calculate the QPC transmission phase
\begin{eqnarray}
\label{TP}
\Delta &=& \delta_T-\delta_I\nonumber\\
       &=& -2{\cal U}+\epsilon \ln\left(4{\cal U}\right)-2\epsilon
-arg\left[\Gamma\left(\tfrac{1}{2}+ i\epsilon\right)\right]
\nonumber\\
       &\to& -2{\cal U}+\epsilon \ln\left(4{\cal U}\right) \ .
\end{eqnarray}
Here we take into account that at $|\epsilon|\lesssim 1$ the argument
of the gamma-function is 
$arg\left[\Gamma\left(\tfrac{1}{2}+ i\epsilon\right)\right]
\approx -(C+2\ln 2)\epsilon =-1.9635\epsilon\approx -2\epsilon$,
see Ref.~\cite{Gradstein}, $C$ is the Euler constant.

We already pointed out that the energy independent part of the transmission
phase is sensitive to the unknown behaviour of the potential far from the saddle point.  Therefore, the constant 
part of the transmission phase can be different from that in (\ref{TP}).
However, the energy dependent part of the transmission phase
is not sensitive to the details of the potential. In essence the energy 
dependent part is calculated with logarithmic accuracy, 
$\ln\left(4{\cal U}\right)\gg 1$, the logarithm enhances the energy dependence
of the transmission phase. In an experiment the height of the potential
is approximately equal to the Fermi energy in two-dimensional leads,
therefore we rewrite the transmission phase (\ref{TP}) as
\begin{eqnarray}
\label{TP1}
\Delta =const +\epsilon \ln\left(\frac{4E_F}{\hbar\omega_x}\right) \ .
\end{eqnarray}
Contrary to the expectation in 
Ref.~\cite{kobayashi13} there is a significant linear energy dependence
of the transmission phase without accounting for electron-electron
interactions.

\begin{figure}[htbp]
\vspace{10pt}
\includegraphics[scale=0.25]{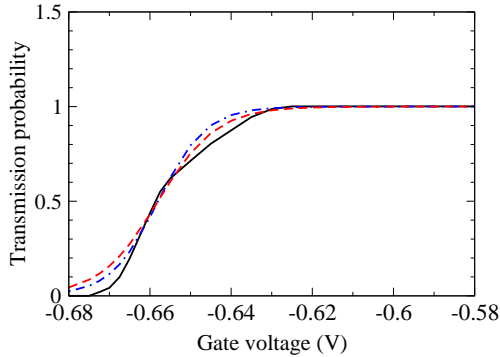}
\caption{Transmission probability versus the gate voltage.
The solid black line shows the data from Ref.~\cite{kobayashi13}.
The red dashed line shows our theoretical fit at $\hbar\omega_x=0.2$meV.
The blue dashed-dotted line shows our theoretical fit at $\hbar\omega_x=0.5$meV.
}
\label{fig2}
\end{figure}

The first attempt to measure the transmission phase was performed by Kobayashi \textit{et al.} in Ref~\cite{kobayashi13}. In their paper they measure the transmission phase of a QPC with two different conductance profiles. After the first measurement the QPC was put through a thermal cycle (heated to room temperature and cooled back down to $200$ mK) before performing the second set of measurements. According to the authors this thermal cycle changes the randomly trapped charged impurities in the 2DEG which in turn changes the conductance profile of the QPC.~\cite{kobayashi13}

In their second set of measurements (after the thermal cycle) the QPC conductance profile exhibits a resonance-like structure in the 0.7 regime ( Fig. 3 in Ref.~\cite{kobayashi13}). This structure is significantly stronger than in all previous measurements of various groups ~\cite{wharam88, vanwees88, thomas96, thomas98, bauer13, micolich11}. It is not clear if the structure observed after the thermal cycle is intrinsic or due to the trapped impurities. Therefore, in this paper, we use only the data from before the thermal cycle (Fig. 2 in Ref.~\cite{kobayashi13}) for comparison with theory.

\begin{figure}[htbp]
\vspace{10pt}
\includegraphics[scale=0.25]{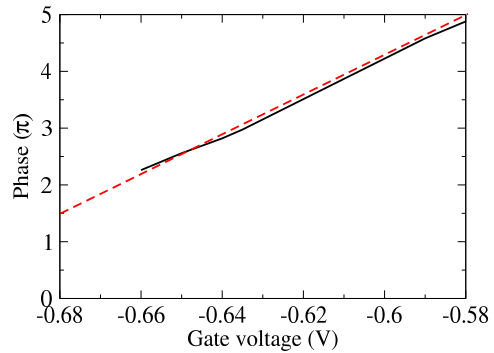}
\caption{Transmission phase versus the gate voltage.
The solid black line shows the data from Ref.~\cite{kobayashi13}.
The red dashed line shows our theoretical fit at $\hbar\omega_x=0.2$meV.
The theoretical fit at $\hbar\omega_x=0.5$meV
coincides with the red dashed line within the line width.
}
\label{fig3}
\end{figure}

In Fig.\ref{fig2} the solid black line is the experimental transmission probability, $ T = \frac{G}{G_0}$ and in Fig.\ref{fig3} the solid black line is the corresponding experimental transmission phase (units of $\pi$) from Ref.~\cite{kobayashi13}.
\linebreak
The Fermi energy in the experiment is $E_F=7.8$meV~\cite{kobayashi13}.
The value of $\omega_x$ is not known, but for all QPCs it is typically
a fraction of meV. We perform fits for $\hbar\omega_x=0.2$meV and for
$\hbar\omega_x=0.5$meV. Fortunately the precise value of $\omega_x$ is
not very important since it appears only under logarithm in Eq.(\ref{TP1}).
The experimental probability and the phase in Figs.\ref{fig2},\ref{fig3} are
given versus the gate voltage, while theoretically these quantities
are calculated as functions of $\epsilon=(E-U_0)/\hbar\omega_x$, 
see Eqs. (\ref{tr}) and (\ref{TP1}).
The only fitting we need to perform concerns a relation between
$\epsilon$ (dimensionless) and the gate voltage $V$ (electron volts).
Naturally we assume a linear relation
\begin{equation}
\label{fit}
\epsilon=\frac{V-V_0}{\alpha}.
\end{equation}
The value $V_0=-0.658$V immediately follows from Fig.~\ref{fig2}.
This is the gate voltage where the transmission is 50\%.
so we are left with only one fitting parameter $\alpha$ to fit
black solid curves in Figs. \ref{fig2},\ref{fig3} using Eqs. (\ref{tr}) and (\ref{TP1}).
The fit with $\hbar\omega_x=0.2$meV gives $\alpha=0.045$,
the fit is shown in Figs.\ref{fig2},\ref{fig3} by the red dashed lines.
The fit with $\hbar\omega_x=0.5$meV gives $\alpha=0.037$,
the probability fit is shown in Fig.\ref{fig2} by the blue dashed-dotted line.
In Fig.~\ref{fig3} the $\alpha=0.037$ theoretical line practically coincides
with the red dashed line.
Overall fits are very good indicating a good agreement between the theory 
and the experiment.

Since the original experimental discovery of the quantized conductance, a 
consistent anomaly at approximately $G = 0.7G_0$ has been noted, commonly 
referred to in literature as the ``0.7 anomaly'' it was first explored 
in Refs.~\cite{thomas96,thomas98} where the authors concluded that the 
anomaly is  due to many body correlation between electrons.
For a recent review of experiments related to the 0.7 anomaly see 
Ref.~\cite{micolich11}.
We believe that the 0.7 anomaly is due to the
enhanced inelastic electron-electron scattering on the top of the potential 
barrier. Analytic theory for this result has been developed in Refs.~\cite{sloggett98, lunde} and functional renormalisation group (FRG) calculations strongly supporting this approach has been performed in Ref.~\cite{bauer13}.
There are also alternative theoretical models of the 0.7 anomaly
based on various assumptions, see e.g. 
Refs.\cite{chuan98,spivak,matveev04,matveev04a,meir02}.

In the present work we do not address the issue of electron correlations in our model.
Nevertheless, we would like to comment briefly how, in our opinion, the correlations can influence the present results. 
It is well known that due to the Coulomb screening the correlations
are not important for higher conductance steps. Therefore, the results are
certainly valid for higher steps. The situation with the first step is 
more complex.
According to the understanding of the 0.7 anomaly
developed in Refs.~\cite{sloggett98,lunde,bauer13} the correlations are
irrelevant at zero temperature. This implies that the results of the present 
work  are valid at $T=0$ even for the lowest step.
At a nonzero temperature inelastic conductance channels are open
and it gives rise to the 0.7 anomaly which scales at $T^2$, 
see Refs.~\cite{sloggett98,lunde,bauer13}.
Due to the unitarity condition an opening of inelastic scattering always influences 
the elastic scattering~\cite{LL}. This implies that the elastic transmission phase
might nontrivially depend on temperature. However, this problem is beyond the scope of the
present work.

In conclusion, using a single-electron picture and the saddle point approximation 
we have calculated the electron transmission phase through a quantum point contact.
Surprisingly,
the transmission phase depends linearly and very significantly on energy of 
the electron.
The phase agrees with recent measurements. Further studies of the transmission phase, both theoretical and experimental (temperature and bias dependence), can shed more light on electron correlations within a QPC and further expand the field. 
We would like to acknowledge important discussions with A. I. Milstein, T. Li and O. Klochan and thank them for their helpful insight.

\end{document}